\documentclass[twocolumn,showpacs,preprintnumbers,amsmath,amssymb]{revtex4}
\usepackage{color}
\usepackage{relsize}
\usepackage{bm, graphicx, amsmath}
\usepackage{hyperref}
\usepackage{float}

\begin{document}

\title{Finding paths in tree graphs with a quantum walk}
\author{Daniel Koch and Mark Hillery}
\affiliation{Department of Physics, Hunter College of the City University of New York, 695 Park Avenue, New York, NY 10065 USA \\ and Physics Program, Graduate Center of the City University of New York, 365 Fifth Avenue, New York, NY 10016}

\begin{abstract}
In this paper, we analyze the potential for new types of searches using the formalism of scattering random walks on Quantum Computers. Given a particular type of graph consisting of nodes and connections, a "Tree Maze", we would like to find a selected final node as quickly as possible, faster than any classical search algorithm. We show that this can be done using a quantum random walk, both exactly through numerical calculations as well as analytically using eigenvectors and eigenvalues of the quantum system.
\end{abstract} 


\maketitle

%
%

\section{Introduction}


\subsection{Quantum Random Walks}
Quantum walks are quantum versions of classical random walks, but because of interference, which is missing in the classical walks, their behavior can be very different \cite{davidovich,aharonov} (for reviews, see \cite{reitzner1,manoucheri}). They have proven useful in a number of algorithmic applications, one of which is searches on graphs \cite{shenvi,potocek,aaronson,lovett,childs,reitzner3,lee,feldman,hillery,hillery2,cottrell,cottrell2}. Initially the searches were for distinguished vertices, that is vertices whose behavior is different than that of the other, normal vertices \cite{shenvi,potocek,aaronson,lovett,childs,reitzner3}. This has since been generalized to searches with non-uniform unmarked edges \cite{lee}, extra edges \cite{feldman,hillery2}, connections between graphs \cite{hillery}, and even a general subgraph \cite{cottrell,cottrell2}. 

A more recent application of quantum walks is in state transfer, where the objective is to use the quantum walk to transfer the particle from a starting vertex to a final vertex, with high probability. It was shown that using a coined quantum walk, a perfect state transfer can be achieved for a star graph and complete graph with self-loops \cite{stefanak}, as well as analysis on a complete bipartite graph \cite{stefanak2}. With the recent experimental realization of discrete-time walks \cite{perets,schmitz,karski,schreiber,peruzzo,schreiber2}, it is hoped that these and other quantum walk applications may some day soon be tested experimentally.

Similar to the premise of state transfers, it has been shown that it is possible to use a quantum walk to find a path between two marked vertices \cite{reitzner2}. In that study, the graph consisted of $M$ linked star graphs. Each star had $N$ edges emanating from a central hub with each edge connected to an additional vertex, which was called an external vertex, to distinguish it from the hub vertex. The stars were arrayed in a line. The first and last stars each had a distinguished vertex, labelled START on the first star and FINISH on the last star. Each star was connected to its neighbor at one vertex, but it was not known which vertex on star $j$, $1 \leq j \leq N-1$, was connected to which vertex on star $j+1$. A classical search would take of order $\mathcal{O}(MN$) steps to find the path between START and FINISH, while the quantum walk only took of order $\mathcal{O}(M\sqrt{N}$) steps.

Here we want to extend the study of finding paths to tree graphs.

\subsection{N-Tree Maze}
The focus of this paper is to explore the possibility of using quantum walks as a means of searching complex mazes that can be represented as nodes and connections on a graph $\mathcal{G}$. To do this, we explore its effectiveness on a specific type of graph, an N-Tree maze.

\begin{figure}[H]
\centering
\includegraphics[scale=0.3]{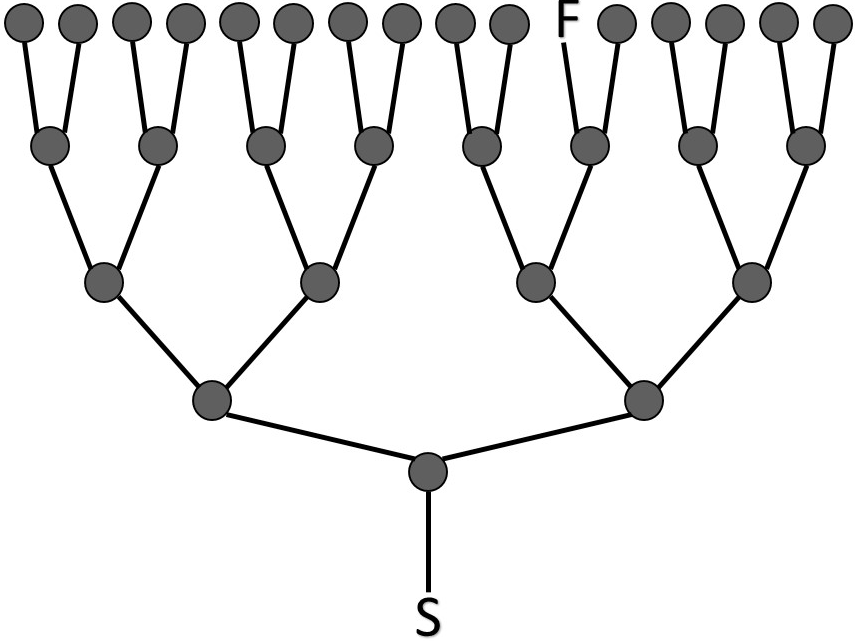}
\caption{An N-Tree Maze, where $N$=2 (also referred to as a 'Binary Tree') and $M$=4. F is always located at an end node at the deepest layer.}
\label{binarytree1}
\end{figure}

An N-Tree maze is illustrated in figure \ref{binarytree1}. From the viewpoint of a solver, the maze consists of junctions with $N$ identical choices. Beyond each choice is another identical junction with $N$ choices, and so on. The maze is characterized as $M$-layers deep, thus to correctly find the exit one must make $M$ correct choices in succession. The figure below would be categorized as $N$=2, $M$=4. If an incorrect path is chosen at any point, it is impossible to know exactly where the mistake was made without backtracking and exhausting all possible paths.

The goal of searching through these graphs is to locate the specified final node F. For this discussion, it is assumed that the location S, the starting node, is known. The location of F is unknown, but $N$ and $M$ are given.

The layout of this paper is as follows. In Section 2, we discuss how a classical computer searches through the maze, and introduce the formalism by which a quantum computer searches using a quantum walk. In Section 3, we discuss the resulting quantum system after implementing the quantum walk, and the way probabilities are distributed throughout the maze. In Section 4, we analyze the various ways one can use the quantum walk to search for F. In particular, we analyze the effectiveness of using the probability concentrated on the edges connected to F, versus using the probability concentrated along the entire path from S to F. In section 5, we derive an expression for the number of trials needed to find F with high probability. And lastly in Section 6, we provide approximate solutions to N-Tree mazes by studying the systems' eigenvalues and eigenstates. We conclude by numerically solving for a form for $\mathcal{U}(N,M)$, the function that gives the number of steps need to prepare a maze of size $N,M$.

%
%

\section{Classical vs Quantum Searches On A Maze}


\subsection{Classical Search Schemes}

For a general maze that can be specified through nodes and connections (also referred to as vertices and edges) on a graph $\mathcal{G}$, often times the best classical search algorithm is a "depth-first" or "breadth-first" search. The two techniques only differ in how they traverse the maze, but share the same underlying principles. The algorithms search recursively through the maze, where one "step" amounts to moving one node at a time and keeping a list of all connected nodes as well as ones previously visited.

If we turn our attention to N-Tree mazes, $M$ layers deep, the search best suited for this maze geometry is a depth-first search. Since the general structure of the maze is known (no internal loops or irregularities) and only the location of F is unknown, the minimum number of steps to reach F is M, while the maximum is $E$, the total number of edges. $E$ is equal to $(N^{M+1}-1)/(N-1)$, thus the average number of steps needed is of the order $\mathcal{O}$ ($ N^{M} $). The real merit of any quantum search algorithm will then be to do better than this scaling.

\begin{figure}[H]
\centering
\includegraphics[scale=0.45]{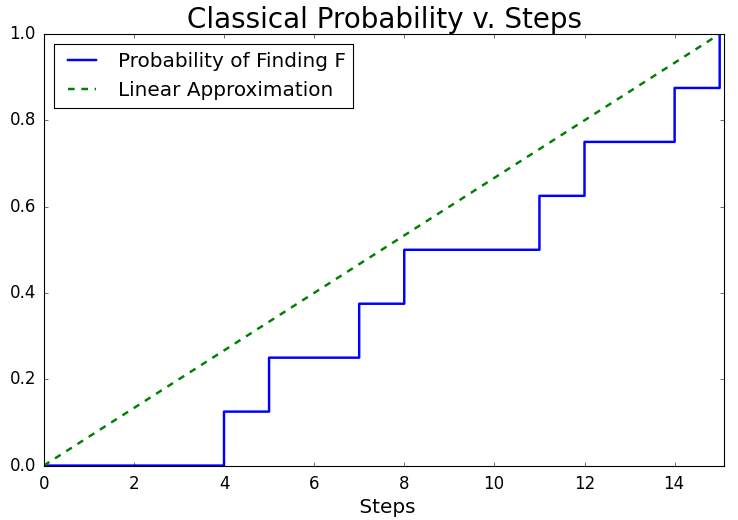}
\caption{The probability of successfully locating F as a function of steps, using a classical depth-first search. Each time an end node is checked, the result is a spike in probability.}
\label{classicalprob}
\end{figure}

Once begun, the classical search will always find F, given enough time. The algorithm moves through the maze and checks all the end nodes in succession, each one with a probability of $\frac{1}{N^{M}}$. Figure \ref{classicalprob} shows the probability of finding the correct final node as a function of steps, searching on a 'Binary Tree' ($N$=2), $M$=3. In this example there are 8 possible end nodes where F could be, thus each jump in the graph corresponds to the algorithm checking one of these nodes.

As the size of these tree mazes get larger, a linear approximation becomes more accurate. We will use these linear approximations to compare classical vs quantum algorithms, and their probabilities of finding the correct final node. On average, the algorithm must check 50 \% of the final nodes before finding the correct one, thus the average number of steps needed is of the order $\mathcal{O}$ ($ N^{M} $). The real merit of any quantum search algorithm will then be to do better than this scaling.


\subsection{Quantum Search using a Random Walk}

To compete with the classical depth-first search, we will use a scattering quantum walk algorithm on the same graph geometry. Using this formalism, the particle resides on the edges of the graph, and can be thought of as being scattered by the nodes. The Hilbert space $\mathcal{H}$ is spanned by $2E$ orthonormal states, two states per edge of the graph. In particular, suppose we have two nodes $A$ and $B$ connected by an edge. Then there is a state $|A,B\rangle$ which represents the particle scattering into node $B$, coming from node $A$, and vice versa for state $|B,A\rangle$.

The scattering walk is a discrete-time quantum walk. The evolution of the system is given by a unitary operator $U$ that advances the walk one time step. This $U$ is obtained by combining the actions of local unitary operations, one for each vertex, which describe the scattering at their respective vertices. In particular, they relate the state entering the vertex at one time step to the state(s) leaving the vertex at the next time step. For a vertex connected to $n$ edges, with $n\geq 3$, the action of $U$ is given by
\begin{equation}
U|j,A\rangle = -r|Aj\rangle + t \sum_{k=1,k\neq j}^{n} |A,k\rangle ,
\end{equation}
where $1\leq j \leq n$, and
\begin{eqnarray}
t &=& \frac{2}{n} \nonumber \\
r &=& \frac{n-2}{n} .
\end{eqnarray} 
The constants $t$ and $r$ can be thought of as transmission and reflection coefficients, respectively. In the special case of end nodes, defined by a node with only 1 connection, $t = 0$ and $r = e^{i\theta}$. For these nodes, $r$ can be any complex number of modulus one, but for our purposes we will only consider the cases where $r$ is equal to either $1$ or $-1$. Specifically, we let the final node F reflect with $-1$, while all other end nodes (including S) reflect with $1$.

Now having defined $\mathcal{H}$ and the evolution of the system, we must choose the initial state of the system. We start with an equal superposition of all states, reflecting the fact that we have no \emph{a priori} knowledge of the location of F. The fact that $U$ acts differently on F drives the quantum system into a nonuniform distribution of probabilities, which is then probed by making a measurement on the system.

%
%

\section{ Quantum Walk on N-Tree Mazes }

In this section we will present results of using a quantum walk on N-Tree mazes. We look at two significant features that arise: 1) The concentration of probability on the correct final node F, and 2) The concentration of probability on the all the states making up the path from S to F. These results are generated numerically. Later we will try to gain a better understanding of them by looking at eigenstates of $U$, also numerically, in section 6.

\begin{figure}[H]
\centering
\includegraphics[scale=0.39]{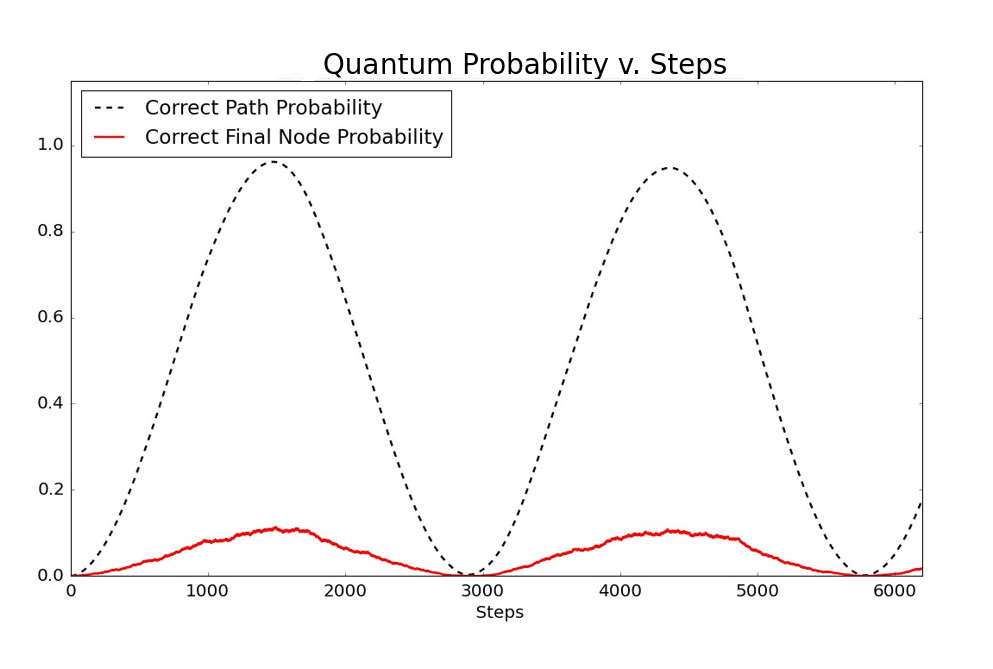}
\caption{Dashed Line: The concentration of probability on states representing the edge connected to F. Solid Line: The sum of the probabilities of the states representing the edges connecting S to F (except the two states directly connected to S)}
\label{probabilityshape}
\end{figure}

In both cases, we find a cyclic rise and fall in probabilities. Figure \ref{probabilityshape} shows an example of these trends.

The location of these peaks (the number of unitary steps needed) increases with maze size, both $N$ and $M$. We discuss this in detail in section 6, but for a rough sense of magnitude, one needs of the order $\mathcal{O}$($N^{\frac{M}{2}}$) unitary steps to reach the peak in probability.


\subsection{ Probability Concentration on F }

By letting the correct final node reflect with -1, and all other final nodes with +1, the result is peaks in probability where F is significantly more probable than any other final node. These peaks come in regular cycles, but only the first peak is ever considered in this discussion. For these first peaks, the table below shows the maximum probability of measuring F for various $N$ and $M$ values.

\begin{figure}[H]
\centering
\includegraphics[scale=0.7]{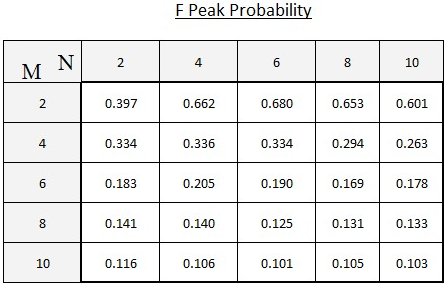}
\caption{Peak Probabilities for measuring a state represented by the edge connected to F.}
\label{Fpeak}
\end{figure}

As shown, the peak probability decreases as $N$ and $M$ get larger, which translates to more trials on average before finding F. However, since this decrease in probability gets smaller as the maze size increases, the effectiveness of using these probabilities for searches will still prove useful.


\subsection{Probability Concentration on the Path}
\begin{figure}[H]
\centering
\includegraphics[scale=0.65]{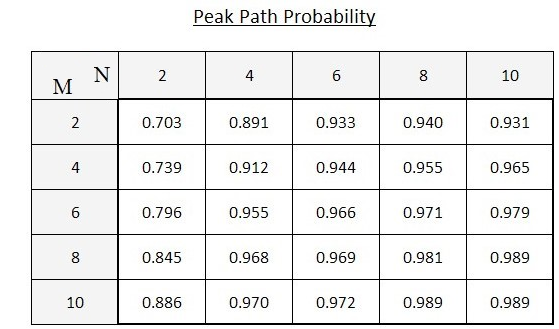}
\caption{Peak probabilities for measuring a state along the correct path, which are states along the path connecting S to F.}
\label{Pathpeak}
\end{figure}

Looking at figure \ref{Fpeak}, we see that at the first peak, the probability that the particle making the walk is on one of the states connected to F is quite low. By contrast, figure \ref{Pathpeak} below reveals that the probability of finding the particle on one of the states connecting S to F (including the states connected to F) is quite high. This result was also shown in \cite{reitzner2}, where using the same type of quantum walk, nearly all of the probability in the system became concentrated along the path connecting a series of stars. For our N-Tree mazes, Figure \ref{Pathpeak} below shows the maximum probability of measuring a state along the correct path for various $N$ and $M$ values.

When we focus our attention to the states connecting S to F, we find that the overall peak probability increases with larger $N$ and $M$. Thus, as the size of the maze increases, it becomes more probable that a state along the correct path is measured. In numerical tests, this trend continued as $N$ and $M$ increase, with the path probability approaching 1 as $N$ and $M$ become very large. 

\begin{figure}[H]
\centering
\includegraphics[scale=0.28]{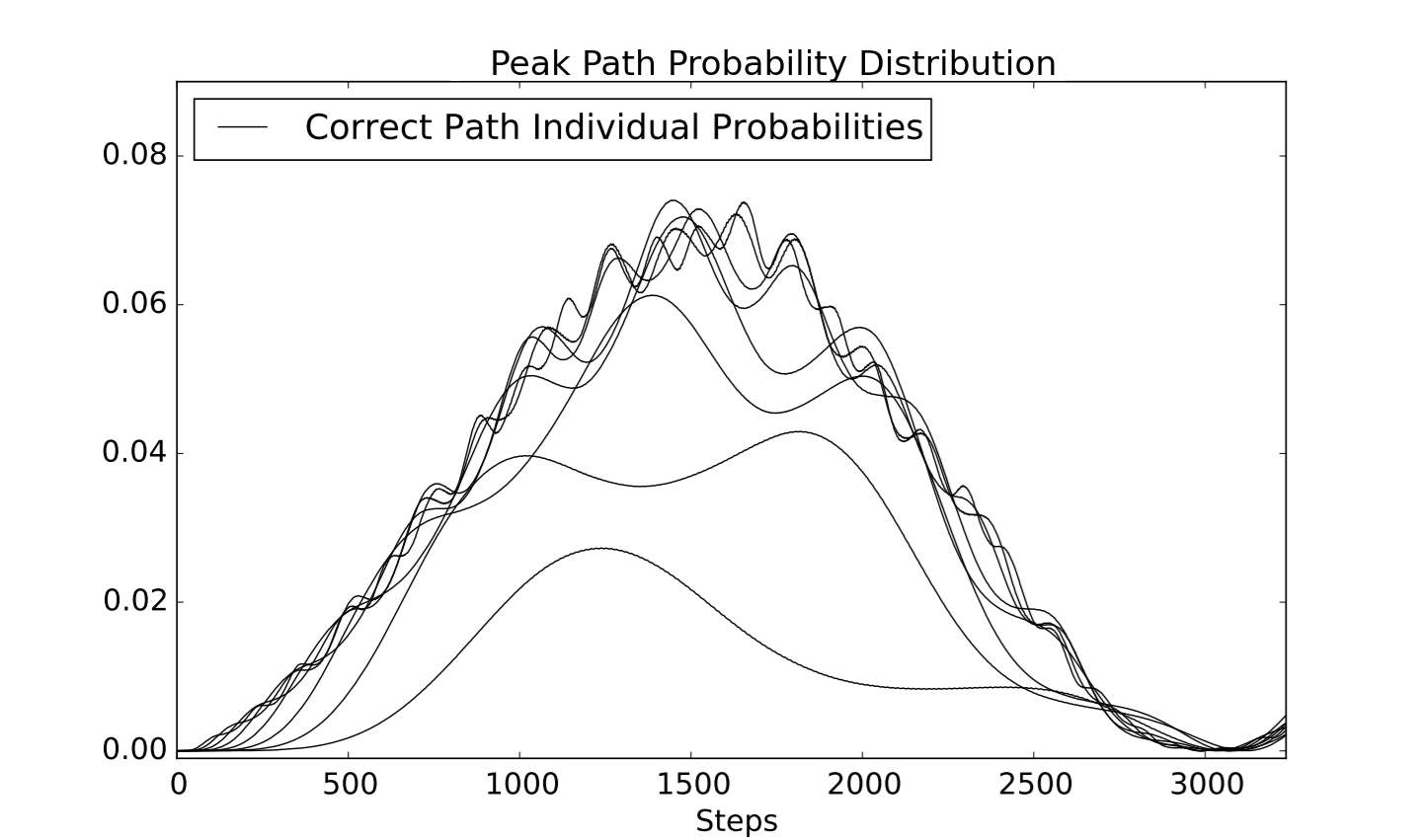}
\caption{Plotted are the probabilities of the 8 closest edges (two states per edge) to F as a function of unitary steps. The states bunched together at the top are closest to F.}
\label{probdistribution}
\end{figure}

If we look at the breakdown of the peak path probability, figure \ref{probdistribution} shows an example of how the individual states each contribute. Most importantly, even with individual fluctuations, all the states along the path peak around the same time, giving rise to the overall shape depicted in figure \ref{probabilityshape}.

The curves that are all very closely clustered together near the top are the states closest to the final node F, while the curves corresponding to lower peak values are closer to S. So when a measurement is made at the moment of the peak path probability, we get a sort of blending of all the correct path state probabilities, slightly favoring states closer to F. This fact will become meaningful in the next section, when we analyze the use of the path probability as a means to search for F.

%
%

\section{Comparing Algorithms and Speeds}

Here we will lay out a few possible algorithm schemes that take advantage of the quantum system, and we compare their effectiveness in finding F. There are two main disadvantages that exclusively plague the quantum algorithms: 1) The system must run a pre-determined number of unitary steps before each measurement. 2) All measurements are probabilistic, so any measurements that fails to find F means that the entire quantum system must be prepared again. We will see that the most effective algorithms work to minimize these disadvantages.

For the quantum algorithm that will come later in this section, we will always be preparing our quantum systems for a peak path measurement. Let us define $\mathcal{U}(N,M)$ as the function that gives us the prescribed number of unitary steps needed to prepare our quantum system for this peak path measurement, for any $N$ and $M$. A complete form fot this function will be given in section 6. For now, we would like to use $\mathcal{U}(N,M)$ as a metric for defining "average speed", which we shall use throughout this paper.

Rather than expressing speed in terms of steps, let us define average speed (for both classical and quantum) as: the ratio of the average number of steps needed to find F to $\mathcal{U}(N,M)$. Using this definition, for a given maze of size $N$ and $M$, the fastest a quantum algorithm can find F is 1. This is because we are always constrained to prepare our quantum system, for a peak measurement, at least once. Thus the theoretical limit of 1 would then correspond to a 100\% success rate of finding F on the first measurement.


\subsection{ Classical Search Speed }

The classical search algorithm works like opening doors, checking each final node to see if it's the correct door, and if not moving onto the next one. Initially, all final nodes have equal probability of being correct, with each wrong node reducing the sample space by 1. Thus, the average speed of the algorithm is determined by the average number of nodes we need to check. 

As depicted in figure \ref{classicalprob}, a linear approximation is sufficient, especially for larger mazes. Equation \ref{p(steps)} gives the linear function corresponding to the probability of success as a function of steps.

\begin{equation}
\label{p(steps)}
P(\textrm{steps}) = \frac{\textrm{steps}}{E} = \frac{\textrm{steps}}{\sum_{i=0}^{M} N^M }
\end{equation}

By setting P(steps) = $\frac{1}{2}$, we get the average number of steps needed for a classical search: Steps$_{avg}$ = $\frac{E}{2}$. Thus, the classical depth-first search algorithm needs to, on average, step through roughly half the maze before finding the correct final node, which scales like $\mathcal{O}(N^{M})$


\subsection{Searching for F Directly}
Suppose we are only interested in searching for the correct final node, discarding all other measurement results. This method is analogous to the Grover search, however, here our chance of success decreases with the size of the maze. As a result, larger mazes will on average expect more failures. Compounded with the fact that larger mazes require more unitary steps to prepare, this type of search suffers drastically from both disadvantages previously mentioned. Nevertheless, we shall see that searching for F directly does indeed provide a speedup over the classical search.

Provided that we know when to probe the quantum system for the maximum probability of measuring F, say given by $\mathcal{U}_{F}(N,M)$, this type of search is analogous to rolling dice. We prepare and probe the quantum system over and over until F is found. If we let p be the probability of measuring F, then our probability of success as a function of trials, $T$, is of the form given in equation \ref{p(t)}. And the average number of trials is given by equation \ref{Tavg}.

\begin{eqnarray}
P_{success}(T) &=& 1 - (1-p)^{T} \label{p(t)} \\
T_{average} &=& \frac{1}{p} \label{Tavg}
\end{eqnarray}

\begin{figure}[H]
\centering
\includegraphics[scale=0.38]{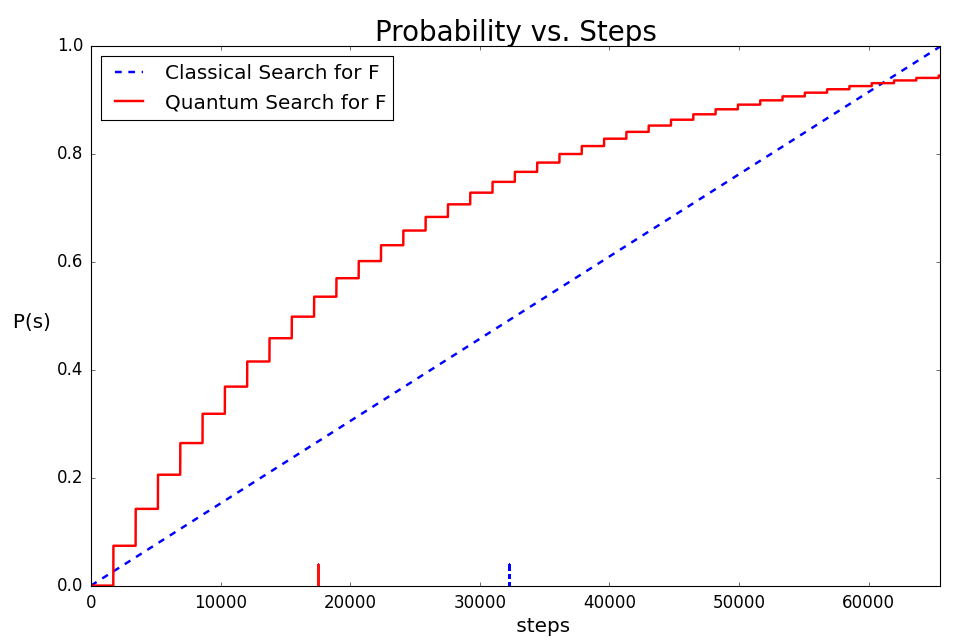}
\caption{Dashed Line: Probability of success as a function of steps for a classical depth-first search. Solid Line: Probability of success as a function of steps using a quantum system to search for F directly. The dashes along the x-axis mark the point for the average number of steps of the two searches respectively}
\label{classicalvsF}
\end{figure}

If we take the number of trials, and multiply by the number of unitary steps it takes to prepare the quantum system, this gives us our relation between success probabilities and steps. Figure \ref{classicalvsF} shows a comparison of these probabilities for the case $N$=2, $M$=15.

Since the two types of probabilities are very different, opening doors versus rolling a dice, one must be cautious in comparing the two. Figure \ref{classicalvsF} suggests that the quantum search is favorable in practically all regions, but when put into practice we want to avoid falling for a Monte Carlo fallacy, thinking the quantum search is faster than it actually is. For this reason, the average number of steps needed to find F for the two searches are marked along the x-axis. In terms of average speed, as defined earlier, we take these average numbers of steps and divide them by $\mathcal{U}_{F}(N,M)$. We then find that the classical search has an average speed of 21.4 while the quantum search is 11.2, thus resulting in a speedup of 1.9. For the quantum search, the average speed can also be viewed as the average number of times the system will need to be prepared.

Figure \ref{Fspeedup} below shows the speedups (ratio of classical to quantum average speeds) for $N$=2, for various $M$.

\begin{figure}[H]
\centering
\includegraphics[scale=0.46]{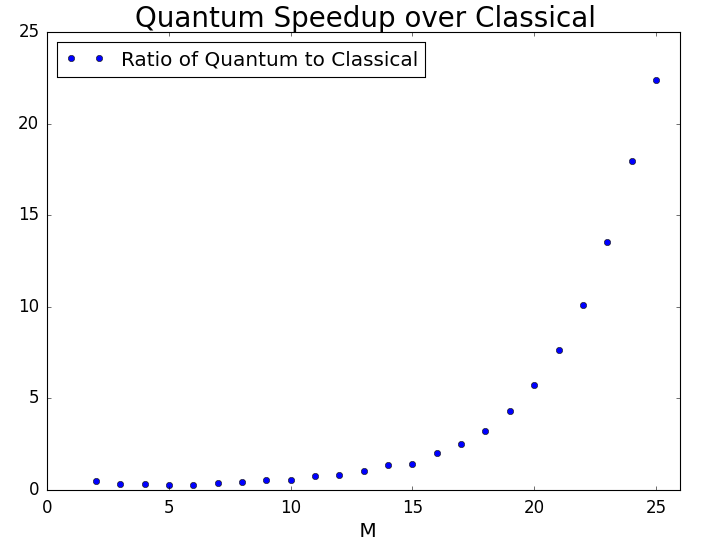}
\caption{Plotted are the speedups of quantum over classical searches, for $N$=2, as a function $M$. These speedups are the result of using the 'search for F directly' algorithm, analogous to a Grover search}
\label{Fspeedup}
\end{figure}

Thus, we do indeed get a speedup by imitating the Grover ideology and searching for F directly. We find that the speedup increases exponentially with the size of the maze. Next, we will show that we can do much better by utilizing what else the quantum system has to offer, specifically the huge peak probability of the correct path.


\subsection{Moving Through The Maze}

The major shortcoming of the pervious search scheme is the amount of wasted steps. Every time a measurement doesn't yield the correct final node, it is discarded. Rather than simply starting over, it would be better if we could extract some kind of meaningful information from these "failed" measurements. 

This is preciously what our next quantum algorithm proposes to do, use the information from previous measurements to guide future ones. To do this, we probe the system when measuring a state along the correct path has the highest probability, rather than maximizing for measuring F. This is because we are anticipating that most measurements will not find F, so instead we will hedge our bets towards the measurement being along the correct path.

\begin{figure}[H]
\centering
\includegraphics[scale=0.40]{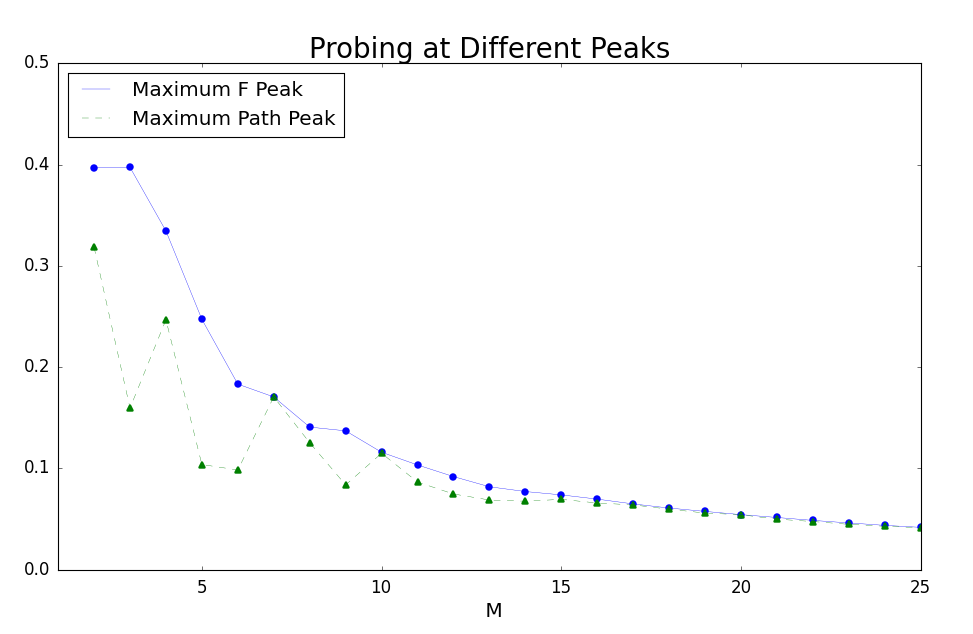}
\caption{Circles: Plotted are the probabilities of measuring F, when the system is prepared to maximize the probability of F. Triangles: Plotted are the probabilities of measuring F, when the system is prepared to maximize the probability of measuring a state along the correct path.}
\label{FatPathpeak}
\end{figure}

Since we will choose to probe at a different time, we need to first make sure that we aren't sacrificing our search for F (which is the ultimate goal) by probing when the path is most probable. Figure \ref{FatPathpeak} shows a comparison of the probability of measuring F, when the system is prepared for maximal F probability versus maximal path probability, for $N$=2. 

At lower $M$'s, particularly 5 and 6, we see a significant decrease in probability for measuring F. However, as the size of the maze increases, we get virtually no loss in probability for F. Thus, for larger mazes, we do not sacrifice anything by probing for a peak path measurement.

Now in order to make use of the following algorithm scheme, it is assumed that we have flexible control over our quantum system. Namely, we have the ability to turn nodes "on" or "off". Specifically, nodes that previously had multiple connections when "on", now act as end nodes with only one connection when "off". Figure \ref{binarytree2} below shows an example. Suppose the edge marked with a star was the result of the first measurement, the node connected to that edge, closer to S, is then turned "off." The result is that all the subsequent nodes behind it are "frozen out," illustrated by the grey dash lines. The remaining edges and nodes form a new graph $\mathcal{G}$ and Hilbert space $\mathcal{H}$.

Changing the geometry of the system is accompanied with a change to the unitary operator U, to properly operate on the new system $\mathcal{H}$. In practice, we will only ever tweak our system at one location, after a measurement has been made. But doing so will always result in freezing out a large section of the initial maze.

\begin{figure}[H]
\centering
\includegraphics[scale=0.24]{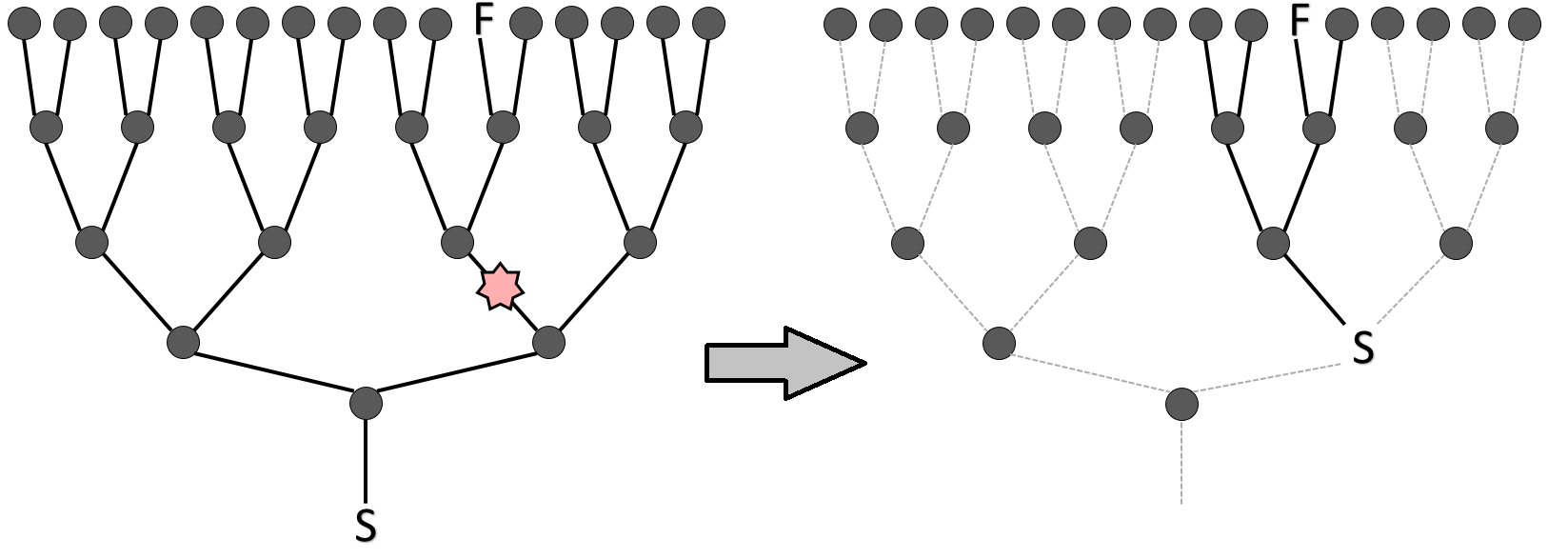}
\caption{Following a measurement (marked by the star), we then turn a single node "off" and consequently freeze out edges and nodes behind it (illustrated as grey-dashed lines). These frozen out edges and nodes no longer affect the active quantum system (illustrated as black-solid lines), and are no longer in the Hilbert space $\mathcal{H}$ or Unitary operator. }
\label{binarytree2}
\end{figure}

Now we are equipped to discuss the core idea proposed by this paper: "moving" through the quantum system. By movement, we mean changing our quantum system after we make a measurement, to reflect the algorithm's scheme for searching through the maze. As shown in figure 10, the result of a measurement dictates which nodes we turn off, or in other words where we "move" through the maze. In example given, the measurement results in a movement from an $M$=4 maze, down to $M$=2. From there, the process is repeated until one of two outcomes occurs: the correct final node F is measured, or an incorrect final node is measured. In the latter case, we must start over from the very beginning, turning on all nodes and preparing the original full maze.

The benefits of moving from an $M$ to a smaller $M$' (where $M$' $<$ $M$) layered maze are as follows: 1) Each preparation of the system will cost fewer unitary steps to reach the peak probability 2) Each measurement will have a higher probability of measuring the correct final node (except for a few cases at lower $M$'s). Thus, if one can successfully move through the maze in increments towards the correct final node, both of the main disadvantages of the quantum algorithm are minimized simultaneously. 

However, the trade-off for movement is the potential new risk of measurement leading to an incorrect step. Any algorithm that utilizes movement through the maze must also have a means of correcting for a movement in a wrong direction, otherwise they could become permanently stuck. We will refer to such an event as stepping into a "dead tree," and analyze its impact on the algorithm next.


\subsection{Moving Into a dead tree}

Let us discuss the event in which a measurement results in a movement in the wrong direction, into a dead tree. Based on the probabilities shown in figure \ref{Pathpeak}, such an even has a small occurrence rate, but is nevertheless worth analyzing. Our interest is how quickly the algorithm finds a final node, and effectively exits the dead tree by resetting the problem back to the initial maze.

When a measurement yields a state that is not along the correct path, the solver has no way of knowing. Thus, following the protocol of the algorithm, one would move to the smaller maze and apply the prescribed number of unitary operations based on $\mathcal{U}(N,M)$. However, since the special node F is no longer in the system, the result of the quantum walk leaves all the states equally probable. But since the structure of each maze is heavily weighted by states closer to the end (refer to figure 1), having all the states be equally probable is a huge advantage for measuring a final node or ones closest to a final node.

\begin{figure}[H]
\centering
\includegraphics[scale=0.39]{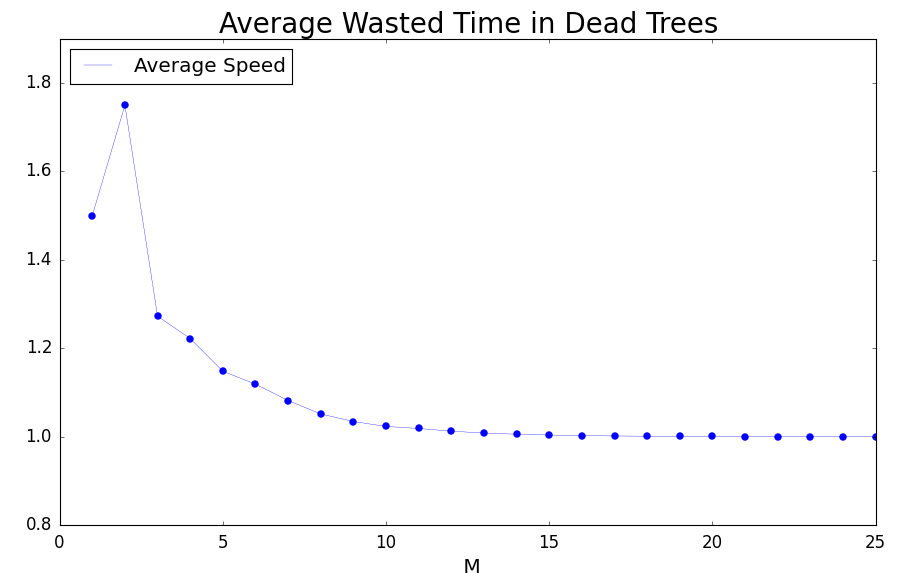}
\caption{The average speed by which the algorithm measures a final node, and effectively exits the dead tree.}
\label{deadspeed}
\end{figure}

As it turns out, the overall loss in speed for misstepping is very negligible. Figure \ref{deadspeed} shows the average speed for which the algorithm exits a dead tree, for $N$=2.

We can see that as $M$ becomes larger, the average number of wasted steps approaches $\mathcal{U}(N,M)$. Because of this, the risk of dead trees slowing down the search algorithm becomes very minimal. In addition, this slowdown of approximately $\mathcal{U}(N,M)$ steps occurs based on the size of the maze \textit{at} the measurement, which in most cases is much smaller than the initial maze size. The largest cost in speed from a misstep actually occurs after the dead tree, when the initial maze, which has the largest $\mathcal{U}(N,M)$, must be prepared again.


\subsection{Results for Following the Measurement Algorithm}

Recall from figure \ref{probdistribution} that states closer to the correct final node, along the path, have slightly larger peak probabilities. Thus, when a state along the path is measured, the resulting layer is on average over half the distance to the final node. After which, each successive measurement does the same. Combine this style of movement with the fact that the maze sizes and $\mathcal{U}$(N,M) scale exponentially, and the result is a powerful searching algorithm.

\begin{figure}[H]
\centering
\includegraphics[scale=0.4]{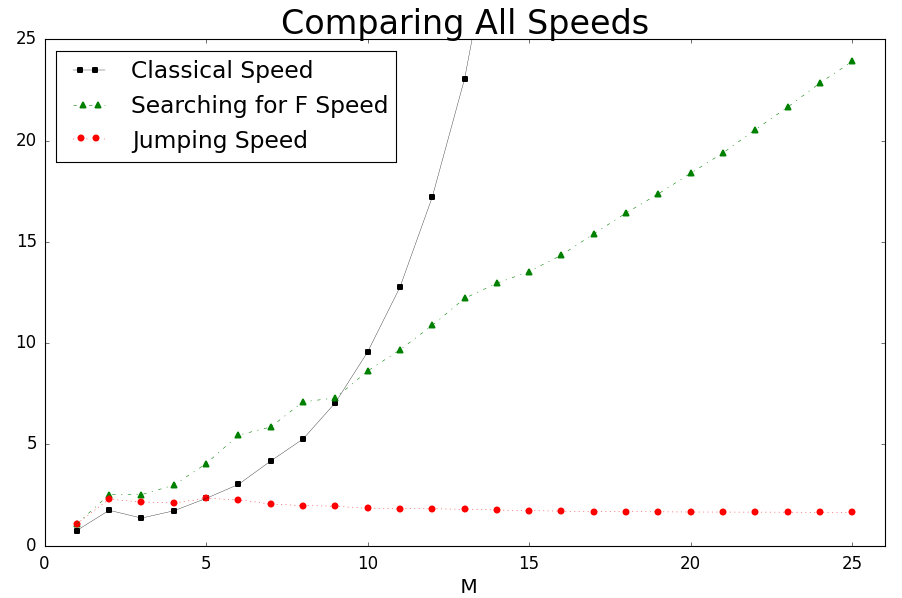}
\caption{Plotted are the average speeds for finding the correct final node F, for all the algorithms discussed: jumping (circles with dashed lines), search for F directly (triangles with dashed lines), classical (squares with solid lines)}
\label{allspeeds}
\end{figure}

Figure \ref{allspeeds} shows the average speed of this algorithm in finding F, as well as the classical and searching for F directly algorithms.

The results show that the movement algorithm outperforms all previous searches. Most notably, as the size of the maze increases, its speed slowly trends toward 1, the theoretical limit. In addition, the fact that the speeds are under 2 means that on average the algorithm finds the correct final node without ever making a misstep. Thus we see the true strength of relying on path probabilities to help guide our quantum searches.

In summary, the quantum algorithm that makes use of movement, and by extension avoids wasting information, outperforms those which don't. In quantum systems where a projective measurement completely collapses the wave function, it is ideal if one can use the information gained from prior experiments to help dictate future ones. For the movement algorithm, the only instances where a measurement yields no information are from measuring S, or measuring a wrong final node.

Whether or not any useful information can be extracted when an incorrect final node is measured, we do not know. We will leave this as an open question for possible future work, and next turn our attention towards some analytical solutions for N-Tree mazes.

%
%

\section{Number of trials}
Let us examine how many trials the movement algorithm will need to find F. In an $N$-Tree maze, which is $M$ layers deep, the correct path from S to F has $M+1$ edges. We will label them with the coordinate $x$, where $1\leq x \leq M+1$. The edge connected to S is $x=1$, $x=2$ is the next edge on the correct path, and $x=M+1$ is the edge connected to F. We shall assume that the probability of the particle being on the correct path is $p$. This probability is close to one, and we will assume that it is independent of the size of the tree, which is a reasonable approximation for all but the smallest trees. We begin by finding the probability of a measurement sequence $x_{1}$, $x_{2}$, $\ldots ,x_{n}$, where $x_{1} \leq x_{2} \leq \ldots x_{n}$, because when we find $x_{j}$, we turn off the tree below this point and only search above it. We assume that each edge on the correct path is equally probable, which our numerical calculations show is a good approximation (In actuality, we know that edges closer to F have slightly higher probabilities). Letting $p_{n}(x_{1},x_{2}, \ldots x_{n})$ be the probability for finding the sequence $x_{1}$, $x_{2}$, $\ldots ,x_{n}$, we find that
\begin{eqnarray}
p_{1}(x_{1}) & = & \frac{p}{M+1} \nonumber \\
p_{2}(x_{1},x_{2}) & = & \frac{p}{M+1} \frac{p}{M-x_{1}+2} ,
\end{eqnarray} 
and in general
\begin{equation}
p_{n}(x_{1},x_{2}, \ldots x_{n}) = \frac{p^{n}}{M+1}\prod_{j=1}^{n-1} \frac{1}{M-x_{j}+2} .
\end{equation}
Now let $P_{succ}(y)$ be the probability of getting the edge connected to F on the $y^{\rm th}$ trial, but not before. We have that
\begin{eqnarray}
P_{succ}(1) & = & \frac{p}{M+1} \nonumber \\
P_{succ}(2) & = & \sum_{x_{1}=1}^{M} p_{2}(x_{1},M+1) \nonumber \\
P_{succ}(y) & = & \sum_{x_{1}=1}^{M} \sum_{x_{2}=x_{1}}^{M} \ldots \sum_{x_{y-1}=x_{y-2}}^{M}
\nonumber \\
& & p_{n}(x_{1},x_{2}, \ldots x_{y-1},M+1) .
\end{eqnarray}
It is possible to approximate the sums in the above expressions by integrals. For example, for $y=3$, we find
\begin{eqnarray}
P_{succ}(3) & = & \frac{p^{3}}{M+1} \sum_{x_{1}=1}^{M} \sum_{x_{2}=x_{1}}^{M} \frac{1}{M-x_{1}+2} \frac{1}{M-x_{2}+2} \nonumber \\
& \simeq & \frac{p^{3}}{M+1} \int_{0}^{M} dx_{1} \int_{x_{1}}^{M} dx_{2} \frac{1}{M-x_{1}+2} 
\nonumber \\
& & \frac{1}{M-x_{2}+2} \nonumber \\
& = & \frac{p^{3}}{M+1} \frac{1}{2} \left[ \ln \left(\frac{M+2}{2}\right)\right]^{2} .
\end{eqnarray}
In general we have that
\begin{equation}
P_{succ}(y) = \frac{1}{M+1} \frac{p^{y}}{(y-1)!} \left[ \ln \left(\frac{M+2}{2}\right)\right]^{y-1} .
\end{equation} 
The probability of finding F on one of the first $z$ trials is
\begin{eqnarray}
\label{trial-find}
P_{find}(z) & = & \sum_{y=1}^{z} P_{succ}(y) \nonumber \\
& = & \frac{p}{M+1} \sum_{y=1}^{z} \frac{p^{y-1}}{(y-1)!} \left[ \ln \left(\frac{M+2}{2}\right)\right]^{y-1} .
\end{eqnarray}

Now consider the sum $f(n,h)=\sum_{k=0}^{N} h^{k}/k!$, for $h>0$. The terms in the sum initially increase as $n$ increases, reaching a maximum when $n$ is the greatest integer less than $h$. After that, they decrease. This suggests that if we choose $n$ to be several times $h$, then the sum will be approximately equal to $e^{h}$. In more detail, using the Stirling approximation, $k!\geq k^{k}e^{-k}$, so
\begin{eqnarray}
\sum_{k=n+1}^{\infty} \frac{h^{k}}{k!} & \leq & \sum_{k=n+1}^{\infty} \left(\frac{he}{k}\right)^{k} 
\nonumber \\
& \leq & \left(\frac{he}{k}\right)^{n+1}\sum_{k=0}^{\infty} \left(\frac{he}{n}\right)^{k} \nonumber \\
& = & \left(\frac{he}{k}\right)^{n+1} \frac{1}{1-(he/n)} ,
\end{eqnarray}
assuming $(he/n)<1$. If we choose $n=rhe$, then the bound is $(1/r)^{rhe}[r/(r-1)]$. So for $he$ of order one, we can choose an $r$ of] order one that will make the above sum small and set $f(n,h)\simeq e^{h}$. Therefore, in Eq.\ (\ref{trial-find}), choosing $z$ to be of the order $p\ln [(M+2)/2]$, we have
\begin{equation}
P_{find} \simeq \frac{p(M+2)}{2(M+1)} .
\end{equation}
Consequently, we will need of the order $p\ln [(M+2)/2]$ trials to find F with high probability.

%
%

\section{Finding Eigenvalues and Eigenvectors}

We have now shown how one can effectively use the path probabilities to find F more quickly (section 4), as well as how the movement algorithm leads to fewer trials need (section 5). Returning to the final point made in the Follow the Measurement section, for larger mazes the average number of steps needed to solve the maze approached the limit of $\mathcal{U}(N,M)$. These results were found numerically, using the exact maximum path probability. But in general for any N-Tree maze, of any size, we would like some form for $\mathcal{U}(N,M)$ that we can use.

In order to do so, we are interested in the eigenvectors and eigenvalues of the operator $U$, which advances the walk one step. This operator acts on a space of dimension $2E$, but we can reduce this by using symmetry. Since we start from an initial state of the system in which all edges have the same amplitude, and we are interested only in later states of the system that can be reached from this initial state by the action of $U$, there will be some edges that always will have the same amplitude. This means that we only need to consider an equal superposition of the corresponding edge states, thereby reducing the dimension of the system. For example, in the tree in Fig.\ref{binarytree1}, the two edges to the left of $F$ will always have the same amplitude, and the four edges to the right of $F$ will have the same amplitude. In both cases, we can replace multiple edges with a single effective edge. Doing so in an N-Tree maze, with $M$ layers, allows us to consider a subspace of dimension $(M+1)(M+2)/2$. From here, we let a computer to find the eigenvectors and eigenvalues of the matrix of $U$ expressed in the basis of effective edge states.


\subsection{Eigenvector and Eigenvalue Characteristics.}

When we let a computer solve for the eigenvalues and eigenvectors, it is found that everything comes in pairs. Specifically:

1) All eigenvalues come in pairs of complex conjugates.

2) All eigenvectors come in pairs, where the values in each vector come in corresponding complex conjugates.

3) $\langle u_{i}^{\ast} | \Psi_{\textrm{initial}} \rangle = \langle u_i | \Psi_{\textrm{initial}} \rangle^{\ast}$, where $|u_{i}\rangle$ is an eigenvector of $U$.

This is a result of the fact that in the basis we are using, the matrix elements of $U$ are real. This implies that the coefficients in the characteristic equation of the matrix are real, hence point 1). For point 2), note that if $U|u_{i}\rangle = \lambda |u_{i}\rangle$, then since $U$ is represented by a real matrix, taking the complex conjugate of both sides gives us that $ |u_{i}^{\ast}\rangle$ is the eigenvector corresponding to $\lambda^{\ast}$. Point 3) follows from the fact that the components of $|\Psi_{init}\rangle$ are real.

Letting $\lambda_{i}$ and $| u_{i} \rangle$ be the eigenvalues and eigenvectors of of the matrix $U$, we can represent the evolution of our system:

\begin{equation}
U^{n}|\Psi_{\textrm{initial}}\rangle = \sum_{i=1}^{d} \beta_{i} (\lambda_{i})^{n} | u_{i} \rangle
\end{equation}
where $\beta_{i}$ is the overlap of the eigenvector with the initial state
\begin{equation}
\label{betadef}
\beta_{i} = \langle u_i | \Psi_{\textrm{initial}} \rangle
\end{equation}

Since the dimensions of the matrix corresponding to $U$ can be quite large as $M$ gets large, one would hope that only a handful of the $\beta$s are dominant, so that smaller terms can be dropped. This indeed turns out to be the case, as all of the $\beta$'s are small enough to be ignored except for two (a corresponding pair of complex conjugates).

Keeping only the largest pair of $\beta$'s and dropping all other terms, our approximate solution becomes:
\begin{equation}
\label{approxeigenform}
U^{n}| \Psi_{\textrm{initial}} \rangle= \beta e^{i\theta_{\lambda} n} |u \rangle + \beta^{*} e^{-i\theta_{\lambda} n} |u^{*} \rangle
\end{equation}
where $e^{i \theta_{\lambda}}$ is equivalent to the eigenvalue $\lambda_i$, expressed in polar form.

Now suppose we are interested in the behavior of a particular state in our original basis. Using the result from equation \ref{approxeigenform}, we can reconstruct a given state, say $|\Phi \rangle $, as follow: 

Let z$_{\phi}=\langle\Phi |u\rangle$, and let W$_{\phi}$($n$) be the amplitude of state $|\Phi \rangle $ after $n$ unitary steps, that is $W_{\phi}(n)=\langle\Phi |U^{n}|\Psi_{initial}\rangle$. Then we have
\begin{equation}
W_{\phi}(n) = \beta e^{i\theta_{\lambda} n} z_{\phi} + \beta^{*} e^{-i\theta_{\lambda} n} z_{\phi}^{*} 
\end{equation}
If we rewrite $\beta$ and z$_{\phi}$ in polar form:
\begin{eqnarray}
&=& |\beta| \cdot |z_{\phi}| \cdot (e^{i(\theta_{\beta} + \theta_{z} + \theta_{\lambda}n)}+ e^{-i(\theta_{\beta} + \theta_{z} + \theta_{\lambda}n)}) \nonumber \\
&=& |\beta| \cdot |z_{\phi}| \cdot 2\textrm{cos}(\theta_{\beta} + \theta_{z} + \theta_{\lambda}n) \label{cosapprox}
\end{eqnarray}
Thus, we get exactly the cyclic form we found numerically. By using the two dominant $\beta$'s, we get the result of equation \ref{cosapprox}, which is that the amplitude of each state behaves sinusoidally as a function of unitary steps. Each edge state can be written as a cosine, with an initial angle of $\theta_{\beta} + \theta_{z}$, that increases by $\theta_{\lambda}n$ after $n$ steps. 

Equation \ref{cosapprox} above is general for all of the states in the system. However, we are primarily interested in the states that make up the correct path from S to F, say $| \alpha \rangle$. For these states, a more appropriate representation will be:
\begin{eqnarray}
| \alpha \rangle &=& |\beta| \cdot |z_{\phi}| \cdot 2\textrm{sin}([\theta_{\beta} + \theta_{z} + \frac{\pi}{2}] + \theta_{\lambda}n) \nonumber \\
& \approx & |\beta| \cdot |z_{\phi}| \cdot 2\textrm{sin}( \theta_{\lambda}n) \label{sinapprox} \\
\nonumber
\end{eqnarray}
This is because for the states along the path, we find numerically that the quantity $[\theta_{\beta} + \theta_{z}]$ is near -$\frac{\pi}{2}$. As a result, the states along the path behave like a sin function, reaching their peak amplitude when $n$ is $\mathcal{O}$($\frac{\pi}{2\theta_{\lambda}}$), offset just slightly by the small initial angle $[\theta_{\beta} + \theta_{z} + \frac{\pi}{2}]$. 

In addition to producing the observed sinusoidal nature, this approximation also tells us that all the states along the path should peak around the same time, which we indeed see. Figure \ref{analyticalapprox} below shows a comparison of the approximation given in equation \ref{sinapprox} versus the true value found numerically, as a function of unitary steps. As we can see, for a relatively small maze, the approximation is very close to the true value. For larger $N$ and $M$ values, the approximation becomes even better.

\begin{figure}[H]
\centering
\includegraphics[scale=0.38]{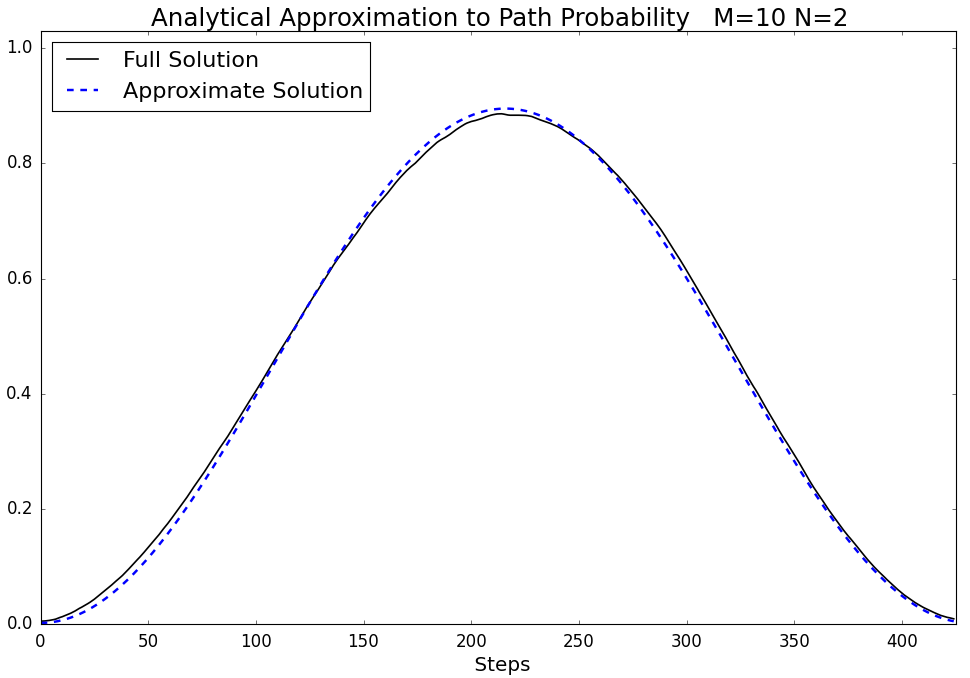}
\caption{Dashed Line: Probability of the path states using the approximation in equation \ref{sinapprox} Solid Line: Actual probability of the path, generated numerically.}
\label{analyticalapprox}
\end{figure}


\subsection{ Extracting Information for Higher Trends }

Using the approximation from the previous section, our real interest is to learn the form of $\mathcal{U}(N,M)$, the function that tells us the required number of unitary steps for a peak path probability. With $\mathcal{U}(N,M)$ in hand, we can extend the search potential of the previously laid out algorithms to any sized maze. 

To get $\mathcal{U}(N,M)$, we are going to use the eigenangles $\theta_{\lambda}$ from the previous section. We will first get the function $\epsilon(N,M)$, which gives the eigenangle $\theta_{\lambda}$ corresponding to the most dominant $\beta$'s, for any $N$ and $M$. With the eigenangles $\epsilon(N,M)$, we can then use the approximation in equation \ref{sinapprox} to find the peak probability.

When a 3D plot is made of the $\epsilon(N,M)$, with $N$ and $M$ on the x and y axes and the value of $\theta_{\lambda}$ on the z, we find an exponentially decreasing function in both $N$ and $M$. Looking at the 3D plot in slices for when we set $M$ to a constant value, we can see recognizable trends for $\theta_{\lambda}$ as a function of just $N$. These trends are of the form y = Ax$^\textrm{B}$, where A and B are constants, and are plotted below in figure \ref{eigenangles}.

\begin{figure}[H]
\centering
\includegraphics[scale=0.48]{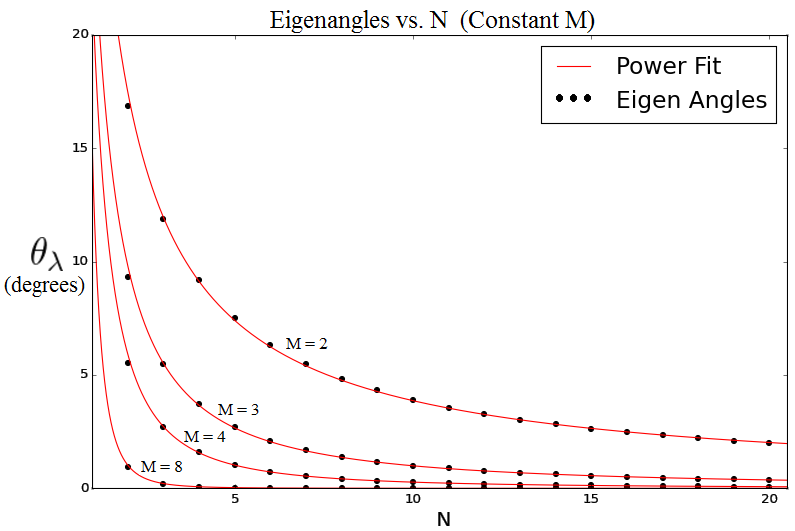}
\caption{Dots: eigenangles $\theta_{\lambda}$, plotted as a function of $N$. Red line: Power fit curve of the form y = Ax$^\textrm{B}$ }
\label{eigenangles}
\end{figure}

This reveals that our function is of the form:

\begin{equation}
\mathlarger{\epsilon}(N,M) = \textrm{A}(M) \cdot N ^ { \textrm{B}(M) }
\end{equation}
Note that $\mathlarger{\epsilon}(N,M)$ here is in degrees. Using the best-fit solutions, shown in figure \ref{eigenangles}, we can analyze the forms for A($M$) and B($M$). Doing so reveals:
\begin{eqnarray}
\textrm{A}(M) &=& \alpha \cdot M ^ {\beta} \nonumber \\
\textrm{B}(M) &=& \rho + \gamma \cdot M
\end{eqnarray}
where $\alpha$, $\beta$, $\rho$, and $\gamma$ are all constants. So we find that A($M$) also takes the form of a power function, while B($M$) is linear.

Putting everything together, we now have our approximate form for $\mathlarger{\epsilon}(N,M)$, and more importantly $\mathcal{U}(N,M)$:
\begin{equation}
\mathcal{U}(N,M) = \frac{90}{\mathlarger{\mathlarger{\epsilon}}(N,M)} = \frac{90}{\alpha} \cdot M^{-\beta} N^{ -\rho - \gamma \cdot M } 
\end{equation}
where
\begin{eqnarray}
\alpha &\approx& 47.87 \hspace{1cm} \beta \approx -.551 \nonumber \\
\rho &\approx& .077 \hspace{1.2cm} \gamma \approx -.498 \label{regressionconstants}
\end{eqnarray}

These constants are all found numerically, using regression fitting. Their exact values may vary slightly by using larger sets of data to generate the regression fits. However, the values provided were generated using $N$ and $M$ up to 15. When tested to see if they reproduce the same peak path probabilities found numerically earlier, they do indeed produce the same peak probabilities, typically within 1-2\% of the exact peak.


\subsection{ Final Remarks }
In conclusion, we have found an approximate form for $\mathcal{U}(N,M)$ through analytical results. Using the regression constants from equation \ref{regressionconstants}, we see that the quantum system peaks around the order $\mathcal{O}(M^{.55} N ^{.5 M}$). Comparing this to the classical speed $\mathcal{O}( N^{M}$), we see a Grover-like speedup. 

But this speedup is only possible by utilizing the "Follow the Measurement" algorithm, which provides us the fastest means of finding F. Specifically, as shown in figure \ref{allspeeds}, the jumping algorithm provides us with an average solving speed that is a small multiple of $\mathcal{U}(N,M)$, roughly 1.5 times for mazes we studied. For larger mazes the $\ln M$ factor from Section V should manifest itself.  However, multiplying $\mathcal{U}(N,M)$ by $\ln M$ provides only an upper bound on the number of steps in the quantum algorithm, because it does not take into account that the size of the trees decreases as we jump up the path.  For this reason, we can truly compare the two solving speeds, and say that the quantum walk provides a definitive speedup over the classical one.  The number of steps in the quantum algorithm has an upper bound of the order $\mathcal{O}(M^{.55} N ^{.5M} \ln M)$. while the number of steps in the classical algorithm is $\mathcal{O}( N ^{M}$).

%
%

\section{Conclusion}

We have developed a modification of the Grover search, where the probability accumulates on a path instead of on a marked vertex. However, when faced with the task of finding a special vertex, hidden in the deepest layer of an N-Tree maze, we have shown that making use of the probabilities of the states leading to the special vertex results in the fastest search. In particular, to capitalize on the system's high density of probability along the correct path, it is necessary incorporate changing the graph, which we called ``movement,'' into the quantum algorithm. The ability to ``mov'' while searching for the special vertex, is analogous to the core element of the classical search.

Our efforts were focused solely on N-Tree mazes due to their high symmetry, and resulting high peak path probabilities. We would like to note that in addition to only F reflecting with -1, we also examined the case where both S and F reflect with -1, but found that letting S be a special vertex results in a less than ideal probability distributions for searches.

It was shown analytically that using the movement algorithm to search for F, one needs on average fewer trials to find F. In addition, by using computational tools to examine the eigenvalues and eigenstates of the various mazes' quantum systems, we provided an equation that reveals the underlying sinusoidal nature of the time-behavior of the system, to very close approximation. And by using these approximations, we were able to produce an approximate form to $\mathcal{U}(N,M)$. Using $\mathcal{U}(N,M)$ along with the average speeds found numerically in section 4, we have ultimately shown that the average solving speed of the quantum algorithm utilizing movement has an upper bound of the order $\mathcal{O}(M^{.55} N ^{.5M}\ln M$), a speedup over the classical speed of $\mathcal{O}( N^{M}$).

\begin{acknowledgements}

This research was supported by a grant from the John Templeton Foundation.

\end{acknowledgements}

\end{document}